\PassOptionsToPackage{numbers,sort&compress}{natbib}  

\documentclass[preprintnumbers,amsmath,amssymb,prd, notitlepage,nofootinbib,twocolumn, superscriptaddress]{revtex4-1}
\usepackage{xcolor}
\usepackage{graphicx}
\usepackage{dcolumn}
\usepackage{bm}
\usepackage{subfigure}
\usepackage{wasysym}
\usepackage{verbatim}
\usepackage{color}
\usepackage{amsmath}
\usepackage[utf8]{inputenc}
\usepackage[normalem]{ulem}

\usepackage[hidelinks]{hyperref}

\newcommand{\be}{\begin{equation}}                                  
\newcommand{\ee}{\end{equation}}                                    
\newcommand{\ba}{\begin{eqnarray}}                                  
\newcommand{\ea}{\end{eqnarray}}                                    
                                          
\newcommand{\barr}{\begin{array}}                                    
\newcommand{\earr}{\end{array}}

\usepackage{aas_macros}
\graphicspath{ {./figures/} }

\begin{document}

\title{Simulated catalogs and maps of radio galaxies at millimeter wavelengths in Websky}
\author{Zack Li}
\affiliation{Department of Astrophysical Sciences, Princeton University, 4 Ivy Lane, Princeton, NJ, USA 08544}
\affiliation{Canadian Institute for Theoretical Astrophysics, University of Toronto, 60 St. George St., Toronto, ON M5S 3H8, Canada}
\author{Giuseppe Puglisi}
\affiliation{Computational Cosmology Center, Lawrence Berkeley National Laboratory, Berkeley, CA 94720, USA }
\affiliation{Space Sciences Laboratory at University of California, 7 Gauss Way, Berkeley, CA 94720 }
\affiliation{Berkeley Center for Cosmological Physics, University of California, Berkeley, CA 94720, USA} 
\author{Mathew S.~Madhavacheril}
\affiliation{Centre for the Universe, Perimeter Institute for Theoretical Physics, Waterloo, ON, Canada N2L 2Y5}
\affiliation{Department of Physics and Astronomy, University of Southern California, Los Angeles, CA, 90007, USA}
\author{Marcelo A.~Alvarez}
\affiliation{Berkeley Center for Cosmological Physics, University of California, Berkeley, CA 94720, USA}
\affiliation{Computational Cosmology Center, Lawrence Berkeley National Laboratory, Berkeley, CA 94720, USA }
\date{\today}

\begin{abstract}
We present simulated millimeter-wavelength maps and catalogs of radio galaxies across the full sky that trace the nonlinear clustering and evolution of dark matter halos from the Websky simulation at $z<4.6$ and $M_{\rm halo}>10^{12} M_{\odot}/h$, and the accompanying  framework for generating a new sample of radio galaxies from any halo catalog of positions, redshifts, and masses. Object fluxes are generated using a hybrid approach that combines (1) existing astrophysical halo models of radio galaxies from the literature to determine the positions and rank-ordering of the observed fluxes  with (2) empirical models from the literature based on fits to the observed distribution of flux densities and (3) spectral indices drawn from an empirically-calibrated frequency-dependent distribution.   The resulting population of radio galaxies is in excellent agreement with the number counts, polarization fractions, and distribution of spectral slopes from the data from observations at millimeter wavelengths from 20-200~GHz, including \emph{Planck}, ALMA, SPT, and ACT. Since the radio galaxies are correlated with the existing cosmic infrared background (CIB), Compton-$y$ (tSZ), and CMB lensing maps from Websky, our model makes new predictions for the cross-correlation power spectra and stacked profiles of radio galaxies and these other components. These simulations will be important for unbiased analysis of a wide variety of observables that are correlated with large-scale structure, such as gravitational lensing and SZ clusters.   
\end{abstract}

\maketitle

\section{Introduction}

The millimeter-wave sky combines a view of the early Universe, through the Cosmic Microwave Background (CMB), with emission and scattering from the full history of extragalactic evolution. Current and upcoming surveys such as \citep[ACT,][]{thornton/etal:2016}, the South Pole Telescope \citep[SPT,][]{benson/etal:2014}, CCAT-prime \citep{CCATPrime}, the Simons Observatory \citep[SO,][]{so_forecast:2019}, and CMB-S4 \citep{cmbs4:2019} involve complex analysis pipelines that include these extragalactic components, and cosmological observables such as lensing convergence, cluster counts, and the Sunyaev-Zeldovich (SZ) effect are correlated with the same large scale structures traced by foreground galaxies. At CMB frequencies, the most challenging foregrounds are the dusty galaxies forming the cosmic infrared background (CIB), especially for the intensity at high Galactic latitudes and at high frequencies, and radio emission from active galactic nuclei (AGN) activity. Radio galaxies are the brightest foreground temperature component at low frequencies (i.e 30-100 GHz), and forthcoming CMB experiments operating at unprecedented sensitivity and sky coverage will be sensitive to clustered diffuse emission from the population of radio galaxies with fluxes $\leq 10$ mJy \citep{dezotti2010}. 

Characterizing these objects is critical because their properties affect the cosmological analysis pipeline significantly, starting with the earliest stages of map-making and mask generation, during which systematic features can be imprinted that could bias results further downstream in the analysis \citep{fabbian2021,lembo2021cmb}. For example, much of the expected fundamental physics payout from future millimeter-wave surveys, such as constraints on the neutrino mass, light relics, and dark matter, come from the damping tail of the lensed CMB power spectrum, the measurements of which are hampered by complexities introduced by radio galaxies. Additionally, with a polarization fractions of a few percent, radio galaxies are a possible foreground for the detection of B-modes from primordial gravitational waves \cite{2005MNRAS.360..935T,2020arXiv200812619T}. Detailed simulations of the radio galaxy population are required to address each of these critical issues.

In this paper, we introduce a model of AGN-powered radio galaxies to the Websky simulation suite \citep{Stein2020}, complementing existing simulations of the CIB, lensing, and the Sunyaev-Zeldovich effect. These simulations will be important for realizing the science goals of current and future surveys in the millimeter, such as ACT, SPT, SO, and CMB-S4. We provide catalogs for these sources for a range of frequencies from 30-300 GHz, relevant for upcoming high-resolution ground-based surveys, as well as providing a package for generating catalogs and maps at other frequencies. The spatial distribution of bright radio galaxies is well-approximated using Poisson statistics and therefore contributes a shot-noise-like term that dominates the power spectrum. We show agreement for source counts throughout the millimeter with all available astrophysical data, and show predictions for cross-correlations of these objects with other components. 
The underlying model for the connection between radio galaxies, separated into the classical Fanaroff-Riley I and II morphology classes \citep{fanaroffriley}, and dark matter halos, is that of \citet{wilman2008} (W08) as implemented and adapted to satisfy constraints from millimeter-wave observations by \citet{sehgal2009} (S10).

Our approach builds on the simulations of S10 by adding a sampling algorithm for fluxes to match observed abundances and spectral slopes, increasing the volume from 1 to 175 (Gpc$/h)^3$, decreasing the minimum halo mass from 10$^{13}M_\odot$ to $1.4\times 10^{12} M_\odot$, and increasing the maximum redshift from $z=3$ to $z=4.6$. The full sky halo catalogs from \citet{Stein2020} we use as an input come from a cubic periodic simulation volume of size $5.6$~Gpc$/h$ and $6144^3$ resolution elements, chosen so that no discontinuities or repetition of structure occur in any direction. More than a decade of new radio source observations in the sub-millimeter wavelength regime from both the ground-based (ACT, SPT, ALMA, ATCA) and satellite experiments (\emph{Planck}) are incorporated through a new empirically-based calibration method for assigning fluxes that explicitly matches observed number counts and object-by-object spectral index measurements while retaining the sophisticated astrophysical halo modeling developed by W08 and S10. As a result, the synthetic catalogs and maps exhibit realistic correlations with other observables connected to halos while simultaneously matching all available data in the sub-millimeter wavelengths.

Similar simulated radio galaxy catalogs have recently been developed for radio surveys at lower frequencies, $f \lesssim 30$ GHz \cite{bonaldi2019, loi2019}. The model implemented in this paper shares many similarities with these models, in particular having a common ancestor in W08. We focus in this paper on higher frequencies from 30 to 300 GHz that are relevant for CMB observations. There has also been recent interest in using machine-learning methods to generate mock CMB foregrounds \cite{Han2021, krachmalnicoff2021}, and the most recent model relevant for radio galaxies \citep{Han2021} results in synthetic two-dimensional maps without redshift catalogs, complementary to our explicitly three-dimensional source catalogs and associated frequency maps. 

We devote Section~\ref{sec:hom} to present the methodology employed to populate the halos with radio sources and   Section~\ref{sec:match} to describe the technique to match the current observations and validate the results.  In Section~\ref{sec:spectra} we show the first predictions from our model for angular correlations between radio galaxies, the CIB and tSZ.  Finally, we discuss and summarize  the results and future outlook in Section~\ref{sec:conclusions}.

\begin{figure}[t]
\includegraphics[width=\columnwidth]{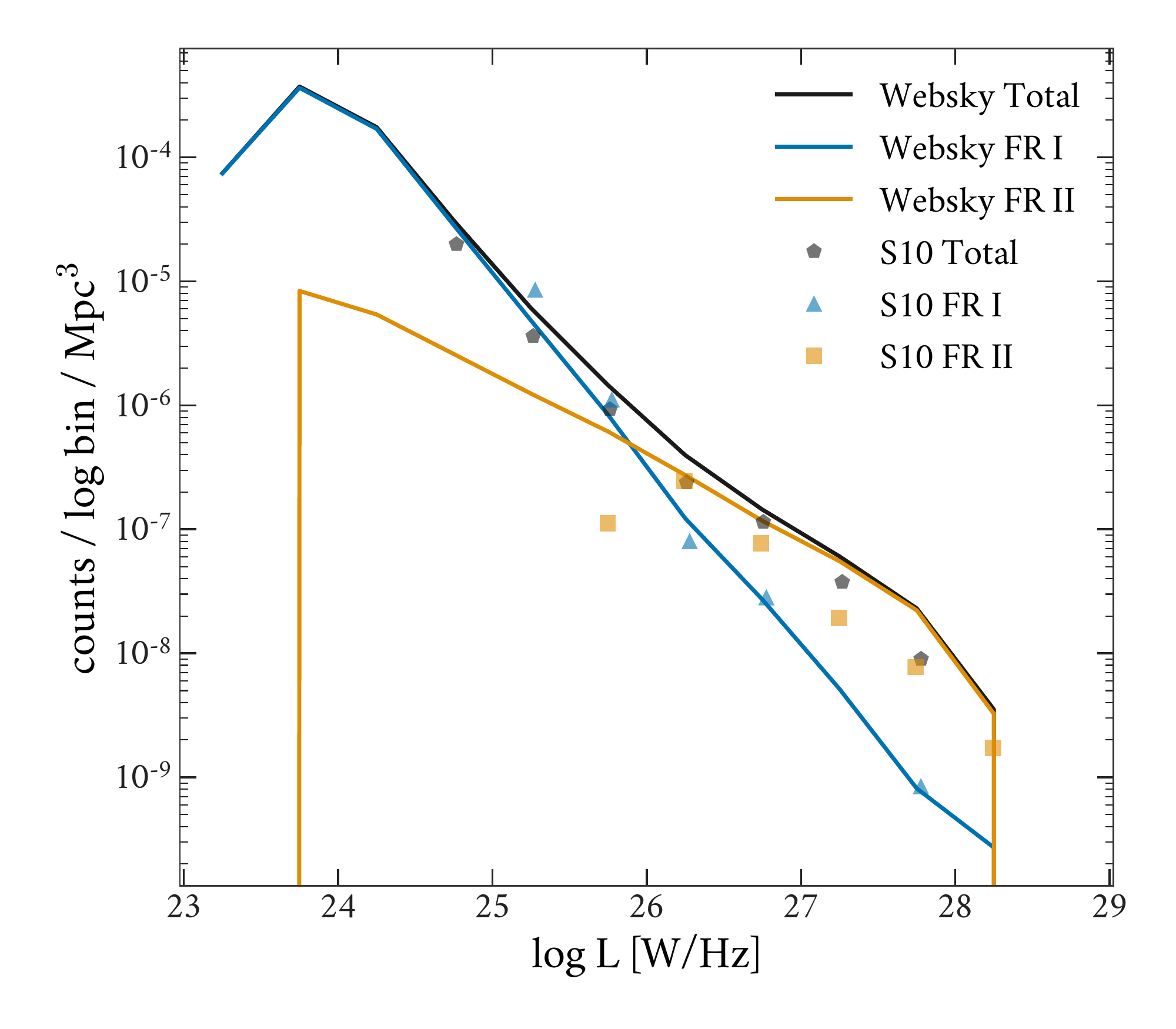}
\centering
\caption{The radio source model used in this paper is anchored on the low frequency (151 Mhz) radio luminosity function, where we expect catalogs to be lobe-dominated and more complete. Here, we show that the Websky counts are consistent with the S10 catalogs at 151 MHz.}
\label{fig:poisson}
\end{figure}

\section{Halo Model}\label{sec:hom}

We implement a model for populating halos with radio galaxies based on S10, itself based on W08. This phenomenological model has shown some disagreement with modern data, as we show in Section III, but we have taken care to preserve the details of this model, for literature comparison, to serve as a baseline for future modeling work, and to aid in untangling the effects of radio sources on cosmology pipelines already tested on S10. Here we summarize this model, and show our reproduction of the counts and spectra from S10 using halos from the Websky halo catalog. The model parameters from S10 are shown in Table~\ref{tab:sehgal_pars}. 
 
The models are constructed to first reproduce the local 151 MHz radio luminosity function (RLF) as in S10, targeting the low frequency RLF which is dominated by steep spectrum, extended sources. We separate radio galaxies into two populations, Fanaroff-Riley Class I (FR-I) and Class II (FR-II). These two populations are observed to have different jet/lobe morphology \citep{fanaroffriley} and very different cosmological evolution. 

\begin{figure*}
     \includegraphics[width=0.95\columnwidth]{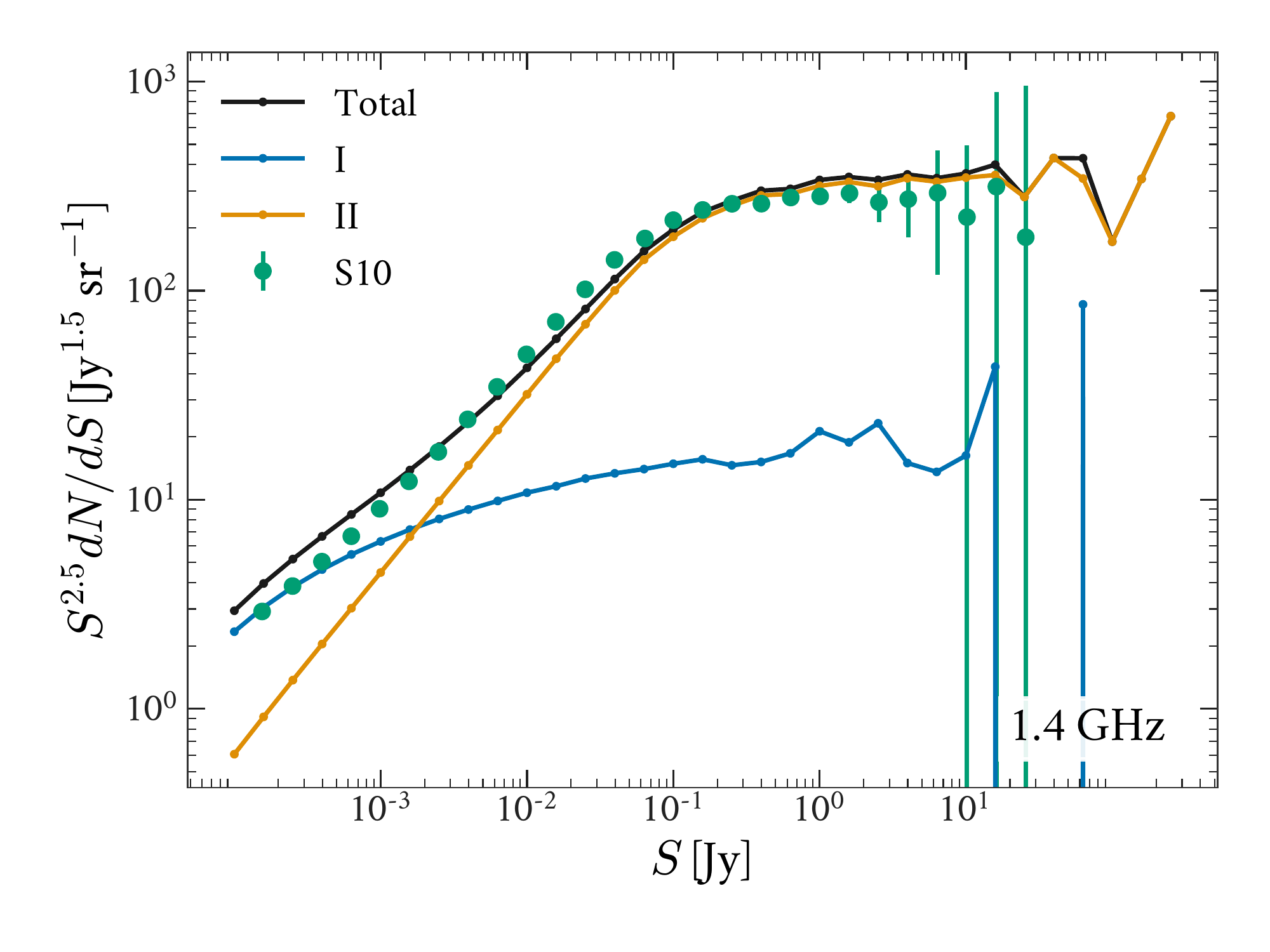} 
    \includegraphics[width=0.95\columnwidth]{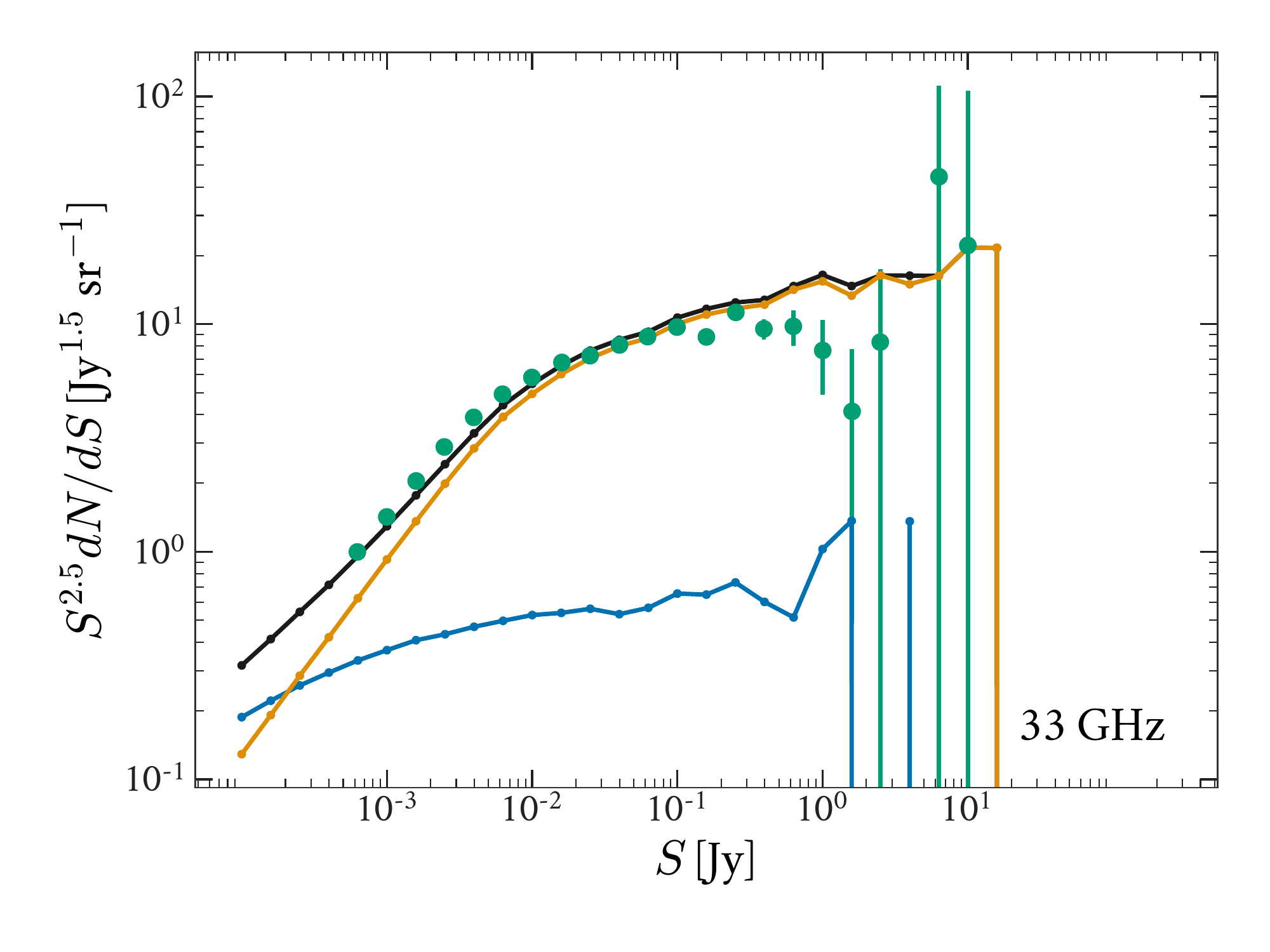}\par
    \includegraphics[width=0.95\columnwidth]{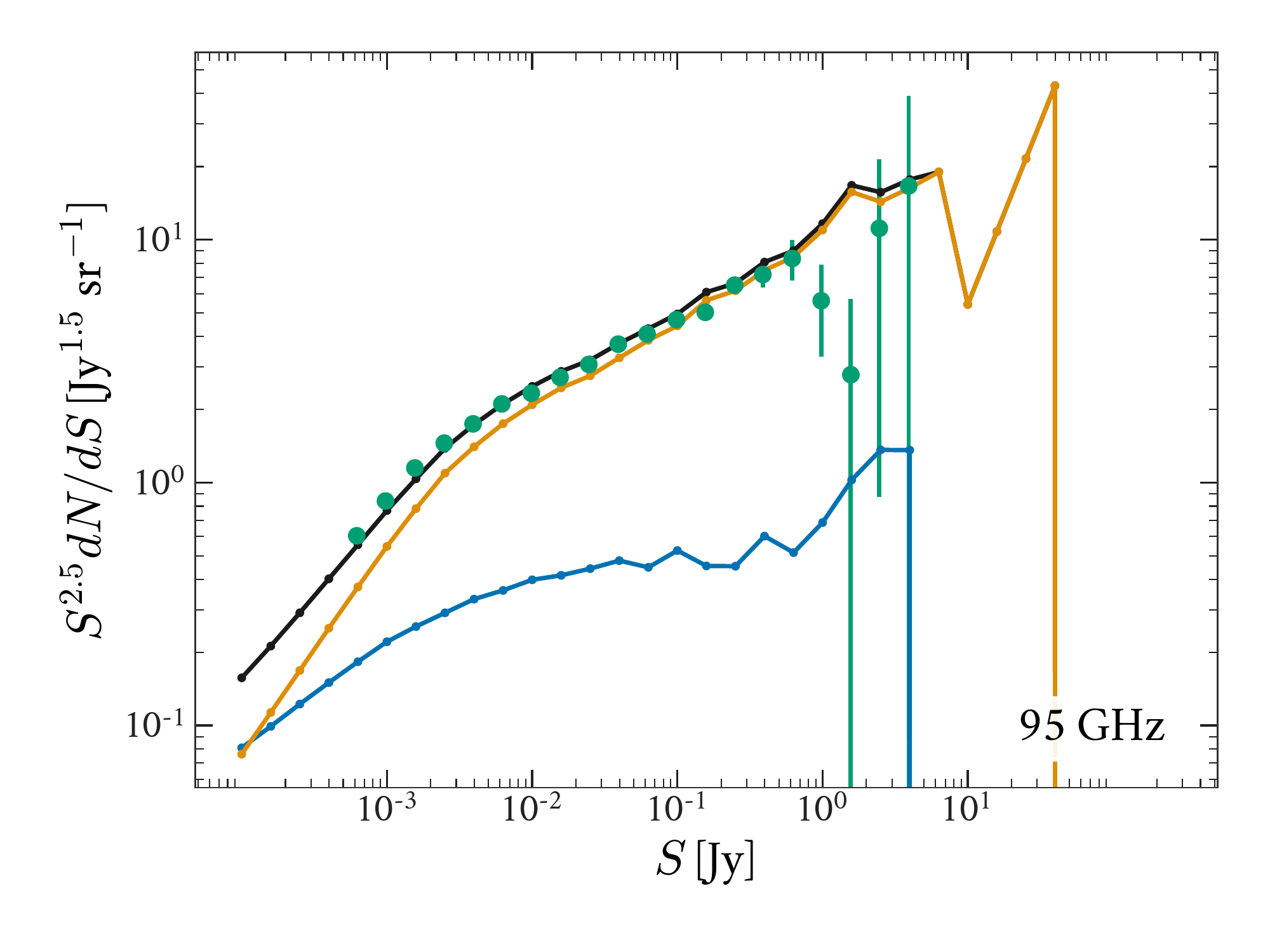}
    \includegraphics[width=0.95\columnwidth]{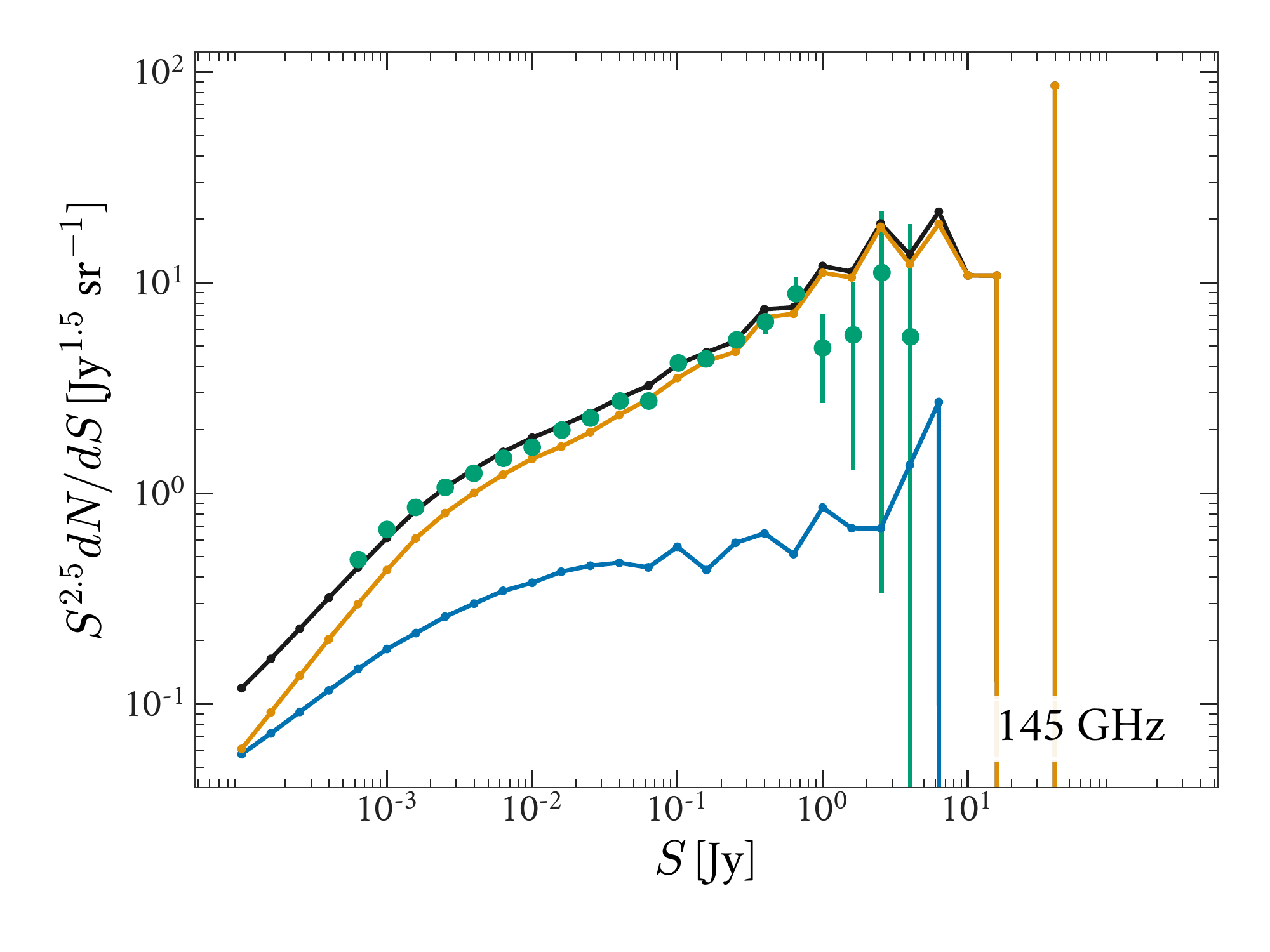}\par
\caption{A comparison of total source counts in S10 (green) and our model (black) at frequencies (from top-to-bottom) 1.4, 33, 95, and 145 GHz. We show the decomposition of the radio source population from our model prior to resampling into FR I (blue) and FR II (orange) morphologies. We provide error bars for the reference S10 catalogs, based on Poisson errors within each bin.}
 \label{fig:sehgal_comparison}
\end{figure*}

\begin{table*}[ht]
\centering
\caption{Model Parameters} 
\label{tab:sehgal_pars}
\begin{tabular*}{\textwidth}{@{\extracolsep{\fill}}*{17}{c}}
\hline \hline 
Type & $R_{\mathrm{int}}$ & $\gamma$ & $a_0$ & $a_1$ & $a_2$ & $a_3$ & $N_0$ & $M_0$ & $\eta$ & $L_b$ & $m$ & $n$ & $\delta$ & $z_p$ & $\sigma_l$ & $\sigma_h$ \\
& & &  & & & & & $(h^{-71}M_{\odot})$ & & W/Hz & & & & & &  \\

\hline
FR I & $10^{-3}$ & $6$ & $0$ & $(-0.12, 0.07)$ & $(-0.34, 0.99)$ & $(-0.75,-0.23)$ & 1 & $4 \times 10^{13}$ & $0.1$ & $10^{24}$ & -1.55 & 0 & 3 & 0.8 & $\cdots$ & $\cdots$ \\
FR II & $10^{-4}$ & $8$ & $0$ & $(-0.12, 0.07)$  & $(-0.34, 0.99)$ & $(-0.75,-0.23)$ & 0.015 & $3 \times 10^{15}$ & 0.1 & $10^{27.5}$ & $-1.6$ & $-0.65$ & $\cdots$ & 1.3 & 0.4 & 0.73 \\

\hline
\end{tabular*} 

\end{table*}%
The mean number of sources per halo, also known as the occupation number, is distributed as 
\begin{equation}
    N(M) = N(z) (M / M_0)^{\eta}.
\end{equation}
with the FR-I population evolving with a broken power law,
\begin{equation}
 N_I(z) = \begin{cases}
N_{I,0} (1 + z)^{\delta} & \text{if }z < z_p \\
N_{I,0} (1 + z_p)^{\delta} & \text{if }z \geq z_p
\end{cases}
\end{equation}
and FR-II evolution modeled as an asymmetric Gaussian,
\begin{equation}
 N_{II}(z) = \begin{cases}
N_{II,0} e^{ - (z - z_p)^2 / (2 \sigma_L^2)} & \text{if }z < z_p \\
N_{II,0} e^{ - (z - z_p)^2 / (2 \sigma_R^2)} & \text{if }z \geq z_p
\end{cases}
\end{equation}
Once $N_I$ and $N_{II}$ are drawn from each halo, we sample a Poisson distribution with parameter $\lambda = N_I$ and $\lambda = N_{II}$ to generate a realization of sources. 

The luminosity function is a broken power law,
\begin{equation}
    p(L) = \begin{cases}  (L/L_b)^m & L > L_b  \\ (L/L_b)^n & \mathrm{otherwise}. \end{cases} 
\end{equation}
This only has finite probably density for $n > -1$ and $m < -1$. Thus, the normalization of this distribution requires a low-luminosity cutoff $L_{\mathrm{min}}$. This is chosen such that the realization of sources correctly translates to the observed 151 MHz luminosity function. As seen in Figure \ref{fig:poisson}, the 151-MHz luminosity function of our catalogs is in good agreement with that obtained from the publicly available catalogs from S10, as expected. 


We separate a radio galaxy into a steep-spectrum extended synchrotron lobe, and a stochastic core hosting a relativistic jet. In this model, these two emission sources dominate in different limits. The steep-spectrum power law emission from the lobes dominate at low frequencies, and their relatively simple emission (optimistically) provides a robust determination of the radio luminosity function. However, the stochastic core emission dominates at higher frequencies. Given a radio luminosity, we impose the luminosity ratio betwen lobe and core, and predict the core emission from the central relativistic jet. This jet is subject to projection effects and beaming, and this cored emission exhibits a flat, stochastic spectral energy distribution that dominates over the lobes at higher frequency. Source counts at CMB frequencies 30 - 150 GHz then provide constraints on the core emission model. 

We begin by drawing random orientations, which are uniformly distributed in $\cos \theta$,
\begin{equation}
    \cos \theta \sim U(0,1).
\end{equation}
We then draw a luminosity from the radio luminosity function described in the previous section. This observed luminosity is subject to beaming, so we follow S10 and term this total, observed luminosity $L_{\text{beam}}$.
\begin{equation}
    L_{\text{beam,151}} \sim p(L).
\end{equation}
This luminosity describes the combined luminosity of the lobe and core components at 151 MHz. We then break this beamed, total luminosity into the intrinsic, unbeamed components. From S10,
\begin{equation}
    L_{\text{beam}} = L_{\text{c}, \text{beam}} + L_{\text{l}, \text{int}},
\end{equation}
since the core emission is beamed from the relativistic jet, unlike the lobe where we observe only the instrinsic emission. We now use the observed core-to-lobe flux ratio to break the luminosity into the core and lobe components. Following W08, we have the relativistic beaming factor 
\begin{equation}
    B(\theta) = \left[ (1 - \beta \cos \theta)^{-2} + (1 + \beta \cos \theta)^{-2} \right]/2
\end{equation}
where $\beta \equiv \sqrt{1-\gamma^{-2}}$ and $\gamma$ is the Lorentz factor of the jet. In this model, $\gamma$ is a parameter of the model, and differs between FR I and FR II objects, due to variation in jet power between the two morphology classes. This factor relates the observed core-to-lobe ratio with the instrinsic, $R_{\text{obs}} = B(\theta) R_{\text{int}}$. We then predict the intrinsic (i.e. before beaming) luminosity,


\begin{equation}
    L_{\text{int}} = L_{\text{beam}} \left(\frac{1 + R_{\text{int}}}{1 + R_{\text{obs}}} \right).
\end{equation}
Given this intrinsic luminosity, we then have our two components of the total observed (beamed) luminosity,
\begin{equation}
    L_{\text{l,int}} = L_{\text{int}} / (1 + R_{\text{int}}),
\end{equation}
\begin{equation}
    L_{\text{c},\text{beam}} = R_{\text{obs}} L_{\text{l,int}}.
\end{equation}
These luminosity quantities have so far been limited to 151 MHz, where the luminosity function was obtained. We now model the lobed emission as a power law in frequency, and model the core emission as a random third-order polynomial in log-luminosity and frequency. 
\begin{equation}
    L_{\text{c,beam}}(\nu) = L_{\text{c,beam,151}} \left[ \sum_{i=1}^{3} a_i \left(\log \frac{\nu}{\text{151 MHz}} \right)^{i} \right],
\end{equation}
\begin{equation}
    L_{\text{l,int}}(\nu) =  L_{\text{l,int,151}} \left( \frac{\nu}{\text{151 MHz}} \right)^{-0.8}.
\end{equation}
The polynomial coefficients $a_i$ are drawn from uniform distributions, with limits provided in Table~\ref{tab:sehgal_pars}. With luminosities in hand, we now can compute a source flux for each object. The differential flux $S_{\nu}$ at redshift $z$ as in \citet{hogg1999}. For comoving distance $D_M$,
\begin{equation} \label{eq:cosmoflux}
    S_{\nu} = \frac{L_{(1+z)\nu} }{4 \pi (1+z) D_M^2} .
\end{equation}
This flux is recorded in the catalog for each frequency, as well as an angular position and redshift, with the source assumed to lie at rest in the center of its host halo. We compare our number counts to those from S10 in Figure \ref{fig:sehgal_comparison} and find good agreement, as expected.
\begin{figure}[t]
\includegraphics[width=\columnwidth]{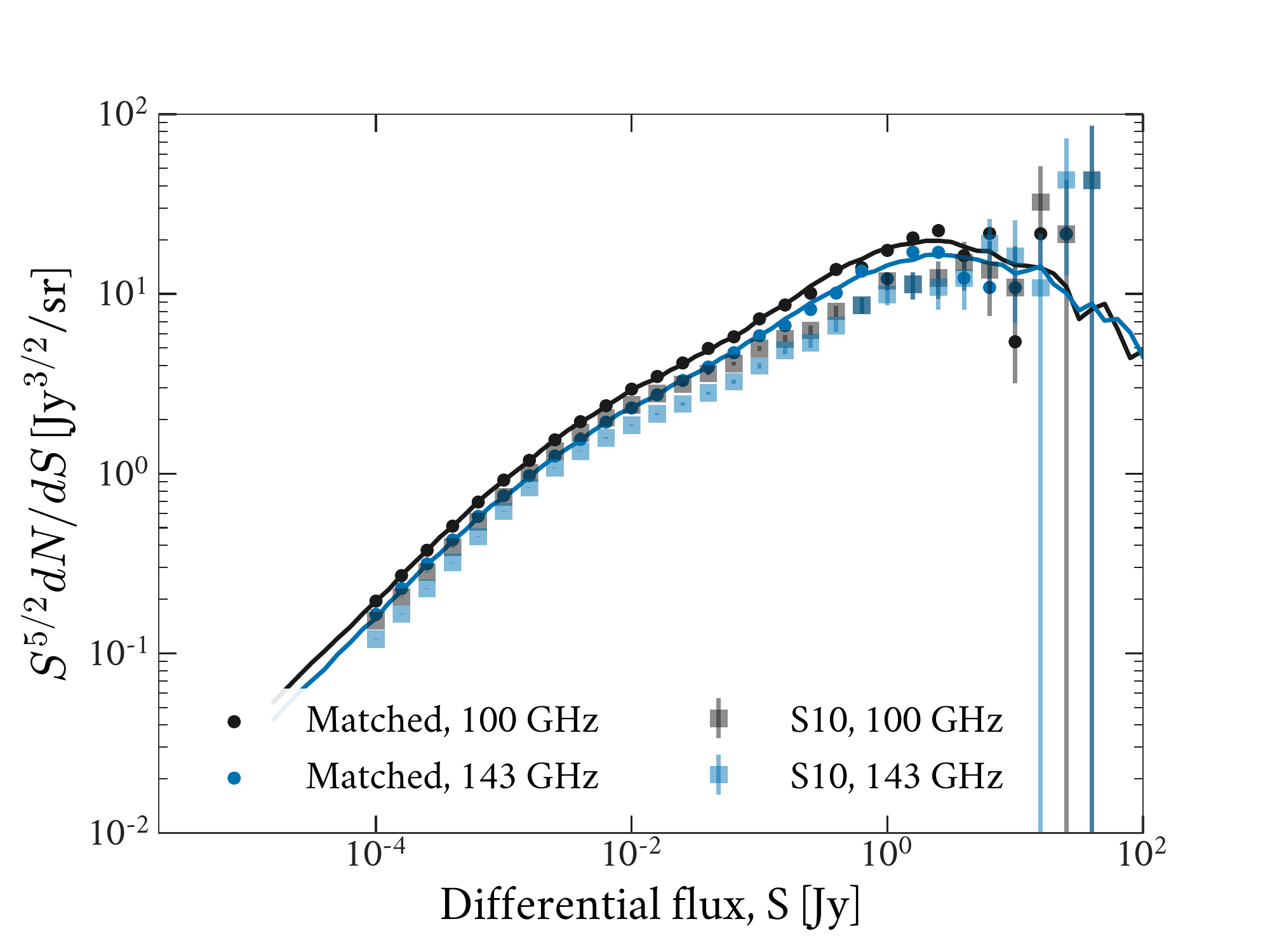}
\centering
\caption{Abundances before (Sehgal, in squares) and after abundance matching (in dots) for 100 and 143 GHz. Error bars are Poisson. The measured counts from Lagache are the solid curves.}
\label{fig:abundance}
\end{figure}

\section{Flux Density and Spectral Index Matching}\label{sec:match}

\begin{figure*}[t]
\includegraphics[width=0.85\columnwidth]{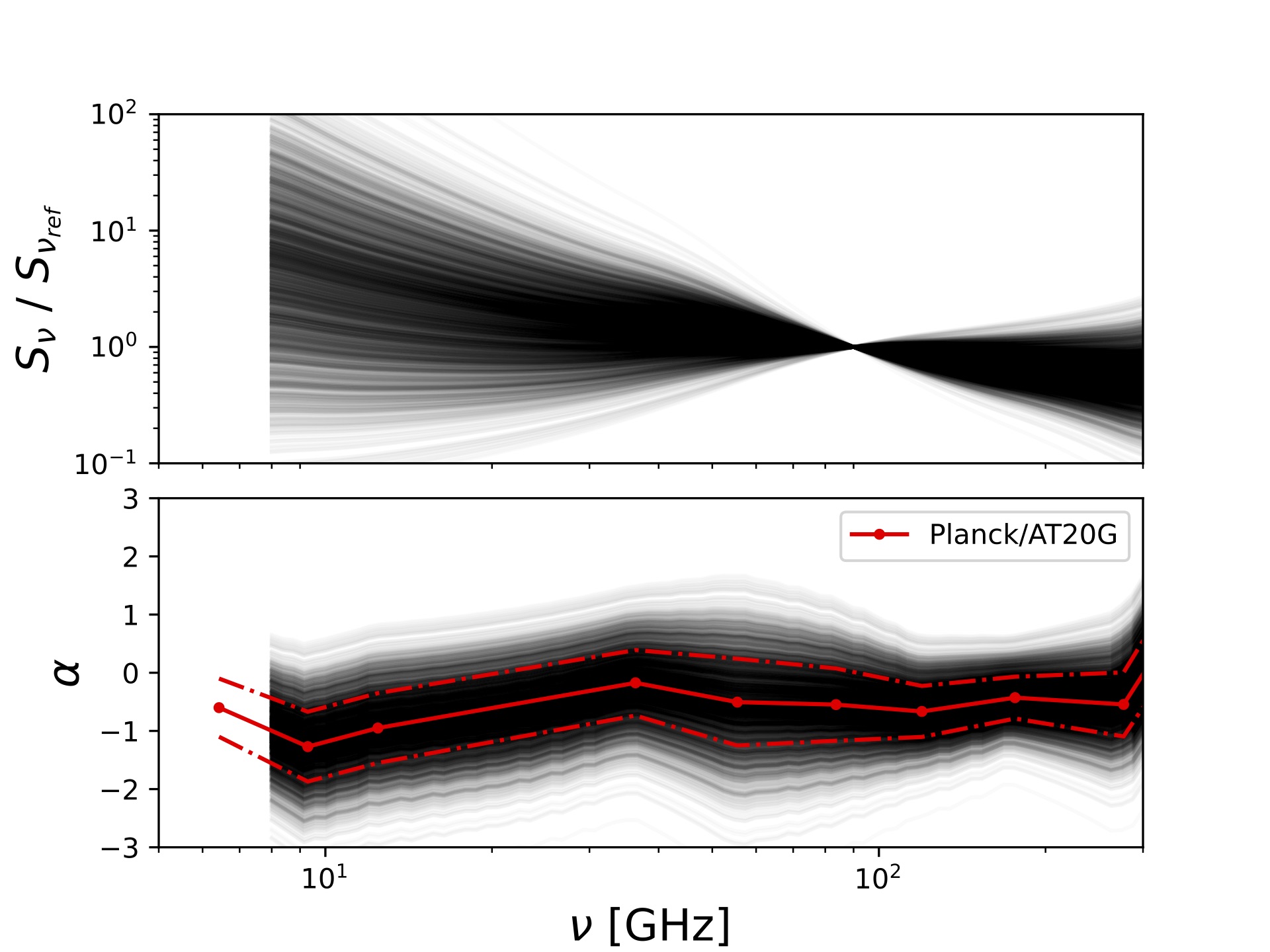}
\includegraphics[trim = 0 15 0 0, clip,width=0.9\columnwidth  ]{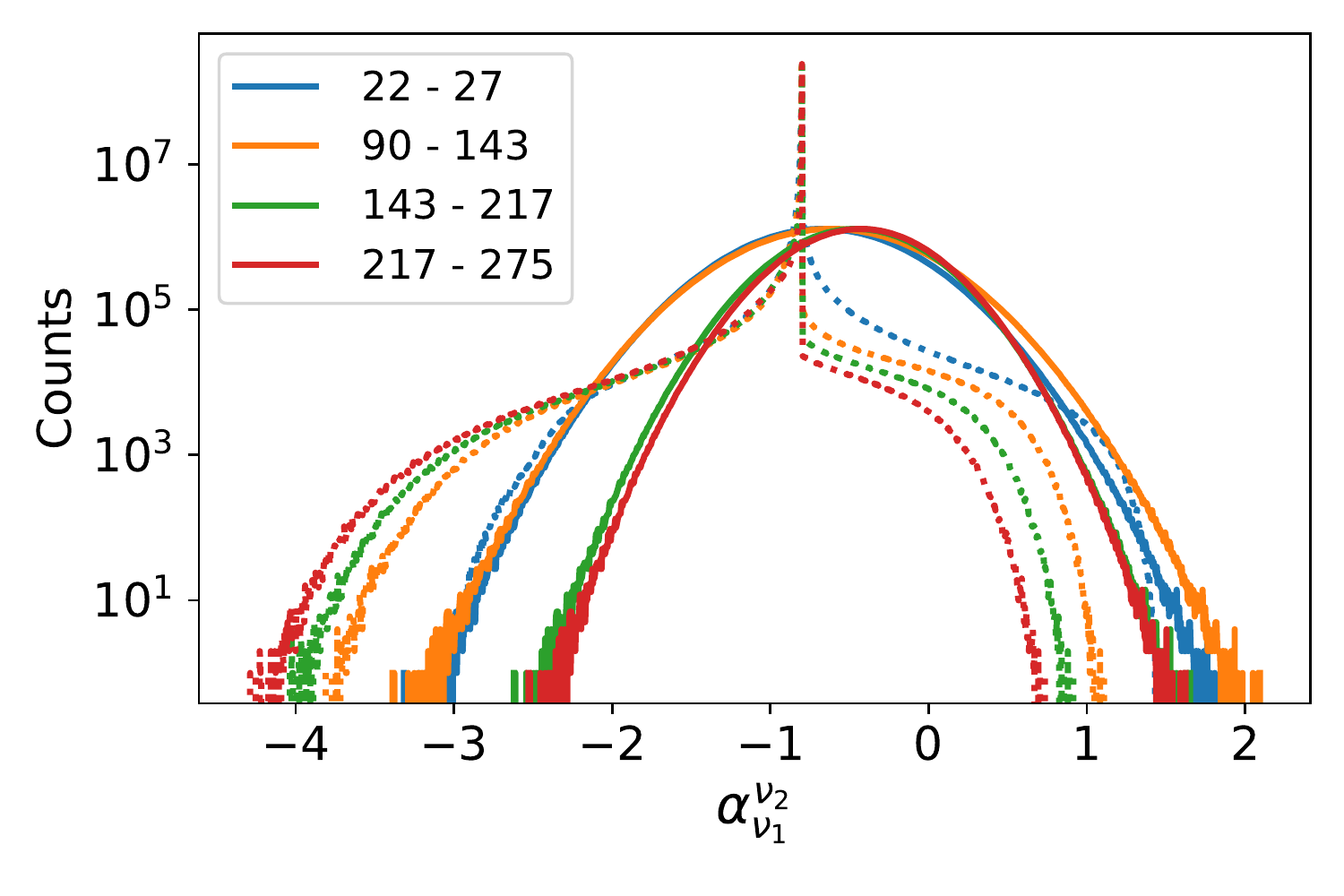}
\centering
\caption{(top left) SED realizations as randomly sampled with the procedure outlined in Section~\ref{sec:match}. (bottom left)  Spectral indices sampled following a Gaussian distribution fit to that measured from the AT20G and \emph{Planck} catalogs. Solid red and dot-dashed lines show the mean and 1-$\sigma$ fluctuations, respectively. (right) Spectral index  distributions for several  choices of  frequencies in GHz, (dashed)  for  fluxes in the catalogs  before matching,(i.e. similar to S10) and (solid) for fluxes in the matched Websky catalogs. }
\label{fig:alpha}
\end{figure*}

As shown in Figure~\ref{fig:abundance},  fluxes estimated with the model in the previous section are not consistent with the latest models of number counts at $S>10$ mJy and at frequencies $>20$ GHz. This is likely because the model in S10 was developed when the most recent radio source data at  sub-mm wavelengths  and at low fluxes were not yet available. 
However, in the last decade data from both space and ground based observations have been used to improve the modelling of radio sources, in particular to characterize the SED of various classes of sources \citep{Massardi2011,DeZotti2015,Galluzzi2017a,Galluzzi2019}, and the number counts \citep{pccs2,Puglisi_2018,Datta2019, Everett2020}.  

To overcome the flux mismatch shown in Fig.~\ref{fig:abundance} due to our halo model, we have implemented a new procedure to generate fluxes at a given set of frequencies in such a way that the output catalog is consistent both with observed flux counts at all frequencies and with the observed SEDs at all frequencies, while still retaining the physically consistent relationship between radio galaxies and large scale structure that is built into the W08 halo model.

As a reference model for the number counts, $n(S,\nu)\equiv dN/dS $, we employ the one  from \citet{Lagache2020} (L20), which is an updated version  of the \citet{Tucci_2011} models given  the  latest data from SPTpol and ACTpol (see L20 for further details). Counts are estimated over a very wide range of frequencies between 5 and 900 GHz. To model the radio source emission at low frequencies ($\nu<5  $ GHz), we use number counts from \citet{2005A&A...431..893D} that have been shown to be in agreement at $<70 $ GHz with the \citet{Tucci_2011} models (see \citet{planck_ercsc,Puglisi_2018}). 

We model the SED by combining the multi-frequency data from \citet{Massardi2011} ( AT20G survey from 5 to 20 GHz) and from \citet{pccs2}  (PCCS2 catalog  from 30 to 217 GHz). The inclusion of AT20G allows us to include in our SED modelling the population of steep-spectrum radio sources whose fluxes are lower than the detection limit ($<10$ mJy) for the \emph{Planck} frequencies. At high fluxes, the blazar population (including BL Lacs and flat-spectrum quasars) are instead very well represented in the PCCS2 catalogs (see \citet{pccs2}).  We can therefore derive $p(\alpha|S,\nu)$, i.e. the distribution of spectral indices at a given frequency and flux by combining the observed distribution of spectral indices, $\alpha(\nu)\equiv \partial{\ln{S_\nu}}/\partial{\ln\nu}$, from AT20G and PCCS2 data. In this work, we assume the distribution of spectral indices is independent of frequency, so that $p(\alpha|S,\nu)=p(\alpha|\nu)$, and fit the available the distribution to the data with a Gaussian. The mean and rms, $\overline{\alpha}(\nu)$ and $\sigma_\alpha(\nu)$, are specified at a set of frequencies and piece-wise linear in $\ln\nu$ between, shown as solid and dot-dashed lines Figure \ref{fig:alpha}, respectively.

\begin{figure*}
\includegraphics[width=2\columnwidth]{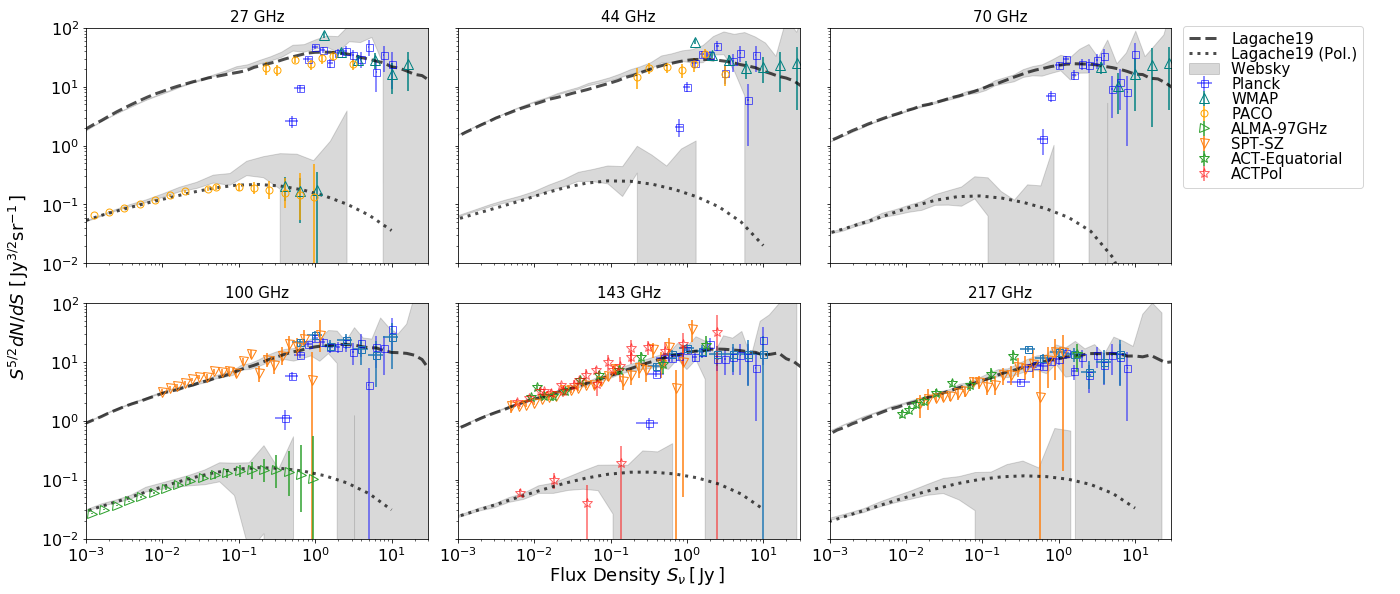}
\centering
\caption{Number counts at several frequency channels. Counts from \citet{Lagache2020} for total density and polarized fluxes respectively in  (black dashed) and (black dotted), (shaded grey area) encode the counts estimated from the mock catalogs described  in this work.   (  upper triangles ) data  from WMAP survey \citep{2009ApJS..180..283W,2010MNRAS.401.1388J}, (  squares) from \emph{Planck} PCCS2 \citep{PlanckearlyXII} and PCCS\citep{PlanckIntermVII},  (circles) counts from \emph{Planck} and ATCA COeval survey \citep{2016MNRAS.455.3249M,Massardi2008,Galluzzi2017b}, (right triangles) counts from ALMA \citet{Galluzzi2019}, (lower triangles) counts from SPT-SZ \citet{Everett2020}, (stars) counts from (green) \citet{Gralla2019} (red) \citet{Datta2019}. }
\label{fig:stateofar_counts}
\end{figure*}

Given the abundance-matched fluxes of each radio object at $\nu_{\rm ref}$, and  $p(\alpha|S,\nu)$, the change in abundance with respect to frequency at fixed flux is given by
\begin{equation}
    \frac{\partial n(S,\nu)}{\partial\ln\nu} = \frac{\partial\left[ n(S,\nu)\overline{\alpha}(S,\nu)\right]}{\partial{\ln S}},
\end{equation}
where
\begin{equation}
    \overline{\alpha}(S,\nu)=\int \alpha p(\alpha|S,\nu)d\alpha. \label{eq:sed}
\end{equation}
This implies that if one specifies $\overline{\alpha}(S,\nu)$ and the number counts at some frequency, $n(S,\nu_{ref} )$, this determines the counts at all other frequencies. 

The abundance matching approach we use to correct the source fluxes from the halo occupation model, to be consistent with observed number counts, follows these steps:

\begin{enumerate}
    \item Select a reference frequency, $\nu_{\rm ref}$, at which to match the number counts, $n(S,\nu_{\rm ref})$, as reported by L20. We use a value of $\nu_{\rm ref}=90 $ GHz for generating all catalogs described in this paper. 
    \item Given an input catalog containing $N_{\rm src}$ galaxies with initial fluxes assigned using the procedure outlined in Section~\ref{sec:hom}, sample $N_{\rm src}$ fluxes from $n(S,\nu_{\rm ref})$ and assign these to the galaxies such that the rank-ordering of the galaxies by flux remains the same. This is the ``abundance-matched" catalog at $\nu_{\rm ref}$.
    \item Assign an SED shape to each galaxy such that the deviation of the spectral slope from the mean, in units of the rms deviation, is independent of frequency, by setting $\alpha(\nu)=\overline{\alpha}(\nu)+r\sigma_\alpha(\nu)$, where $\overline{\alpha}(\nu)$ and  $\sigma_\alpha(\nu)$ are modeled as described above and shown in Figure 4, and the frequency-independent value $r$ is sampled randomly from a standard normal distribution. 
    \item Polarized fluxes  are generated  by randomly sampling from a lognormal distribution with median fractional polarization  $ \Pi_{med} \sim 2 \% $, consistent with  recent observations  \citep{Puglisi_2018,Datta2019,Gupta2019}.  
\end{enumerate}

In Figure~\ref{fig:stateofar_counts}, we show the resulting number counts as  shaded grey areas (we include the Poissonian uncertainties on each bin of flux) compared with the L20 ones (long dashed). We observe a remarkable agreement not only  between the counts of the L20 and Websky models at the reference frequency, but also with all available data at other frequencies.  We also show the polarization number counts which result from the convolution of the polarization fraction distribution with the total intensity number counts.   
 
\section{Power spectra} \label{sec:spectra}

We now compute power spectra of our simulated maps with the other foregrounds in the Websky simulation suite. These frequency dependent cross-spectra enter directly into the marginalized bandpowers of the primary CMB power spectrum, so are of particular interest. However, we note that these spectra do not capture the full non-Gaussian information available in our simulated maps caused by the nonlinear clustering and other effects associated with galaxy formation captured in our halo modeling.

We generate an apodized mask from our simulated point sources for use in our power spectrum analysis, in order to mimic pipelines for current and upcoming millimeter-wave experiments. We generate a point-source mask from our simulated catalogs with a flux cut of 7 mJy selected at 143 GHz. We fully exclude pixels with sources brighter than the flux cut, and then further apodize the point source mask with a 15 arcminute "C2" apodization \citep{grain2009}. This same mask is then applied to all frequency maps generated from the point source catalogs. 
All auto and cross-spectra we present here are computed on maps with the point source mask applied. 

We then bin the simulated point sources into a HEALPix \citep{HEALPix} map with resolution $n_{side} = 4096$ for each frequency. This binning operation introduces a pixel window function, which is corrected during the mode decoupling process.
We neglect the effects of beams, since a symmetric beam is trivially corrected for in a typical analysis.
To correct for the effect of the mask, we then perform a mode-coupling calculation with NaMaster \citep{namaster}. 

In Figure \ref{fig:spectra143}, we show spectra of radio galaxies, thermal Sunyaev-Zeldovich (tSZ), and cosmic infrared background after applying the point source mask and decoupling the effects of that mask. Given a realistic flux cut for future surveys derived from the radio galaxy catalog, our simulated radio galaxies are subdominant at 143 GHz until $\ell \sim 3000 - 4000$. Despite masking the brightest sources, the auto-spectrum of our radio sources still follow a roughly Poissonian shape as a function of multipole.

\begin{figure}[t]
\includegraphics[width=0.75\columnwidth]{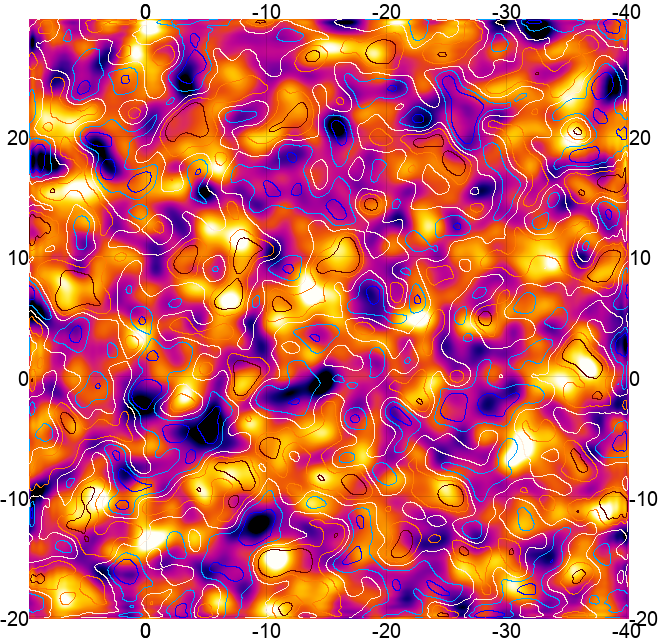}
\centering
\caption{The Websky radio galaxy density (for all sources with flux less than 1 mJy) is shown here in a $50~ \rm{deg.} \times 50~ \rm{deg.}$ field, overlaid with contours denoting the CMB lensing convergence (for $z<4.6$). Both fields have been additionally filtered with a Gaussian beam of FWHM 1.7 degrees. Contours for CMB lensing are spaced by the RMS of the convergence. Since the radio galaxies trace the same underlying dark matter in the CMB lensing field, a clear correlation by eye can be seen between the white overdensities (black voids) in the radio galaxy field and the red peaks (blue valleys) of the CMB lensing field.}
\label{fig:kappacorr}
\end{figure}

In Figure \ref{fig:corrcoeff}, we show the correlation coefficients $\rho^{AB} = C_{\ell}^{AB} / \sqrt{C_{\ell}^{AA} C_{\ell}^{BB}}$ between these extragalactic foreground components. The well-known $\sim$30\% correlation between the tSZ and CIB is shown in blue, and this correlation has become a necessary part of multifrequency power spectrum analysis for CMB experiments \citep{Dunkley2013}. The correlation between radio galaxies and CIB (black), tSZ (yellow), and lensing convergence (green) are all at the $\sim$10\% level. In Figure \ref{fig:kappacorr}, we provide a a visualization of the correlation of gravitational lensing with radio galaxies (with a 1 mJy flux cut to enhance the overall correlation strength). Although the correlation between radio galaxies and other extragalactic foregrounds is only about a third as strong as the correlation between tSZ and CIB, accounting for this correlation may become necessary for multifrequency likelihoods from new millimeter-wave surveys. New, high-resolution data will place increasingly stringent constraints on the clustering of these extragalactic foregrounds.


\begin{figure}[t]
\includegraphics[trim = 0 30 0 0, clip,width=0.9\columnwidth]{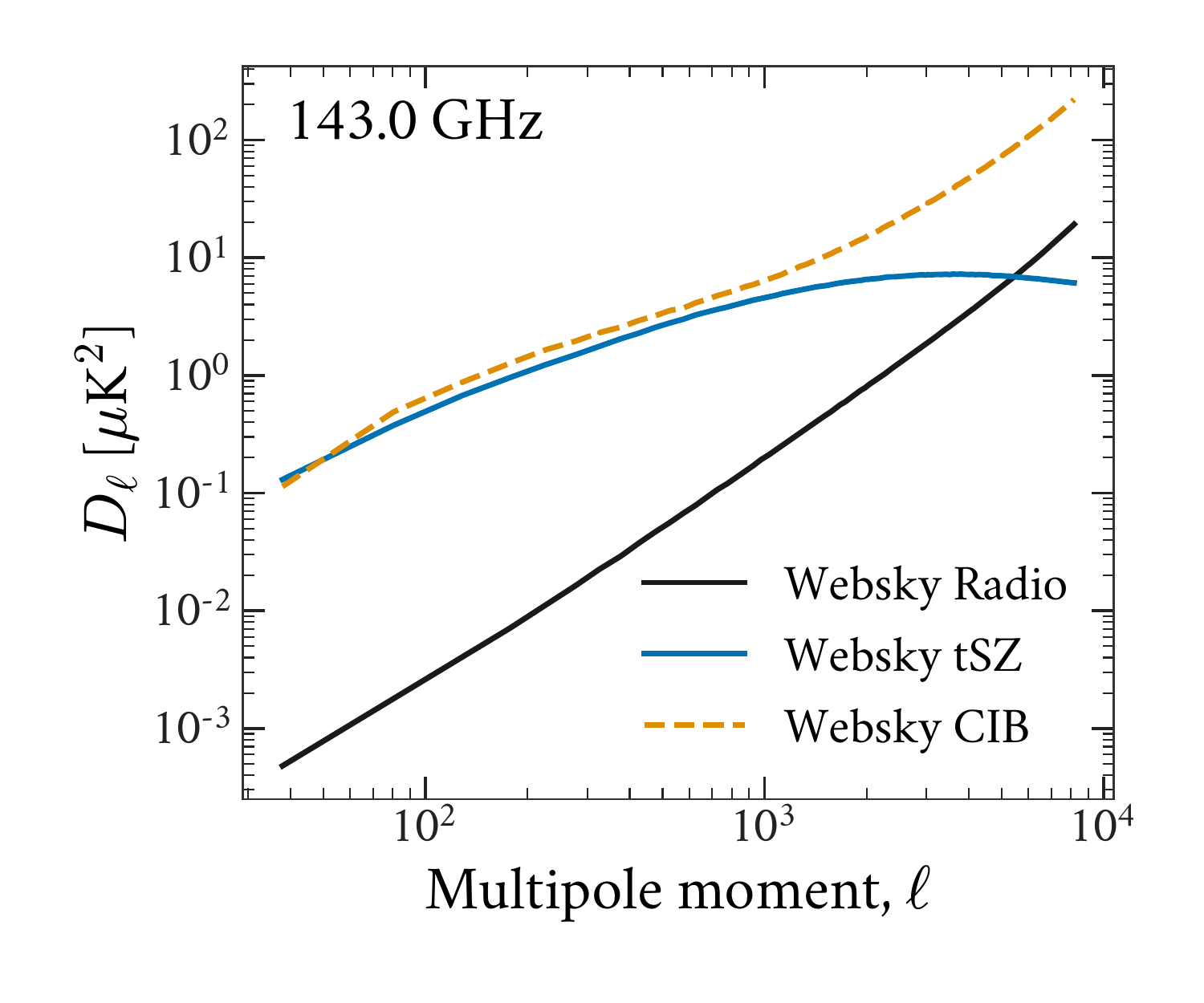}
\centering
\caption{Power spectra at 143 GHz for the Websky radio galaxies (this work) compared to the power spectra generated from thermal Sunyaev-Zeldovich (tSZ) and the Cosmic Infrared Background (CIB). We estimate realistic spectra by generating a sky mask from the brightest radio point sources, and estimate spectra on the masked maps using pseudo-C$_{\ell}$ methods (see Section \ref{sec:spectra}).}
\label{fig:spectra143}
\end{figure}

\section{Products and Code Release}
The catalogs as well as the maps will be made publicly available\footnote{ \url{https://portal.nersc.gov/project/sobs/users/Radio_WebSky/}}. 
We will also release the catalog output of the halo occupation model (Section~\ref{sec:hom}) and the catalogs after matching the observations as described in Section~\ref{sec:match}.  Catalogs are released in a very wide range of frequencies (from 20 -800 GHz). We specifically include the  nominal frequencies for SO, CCAT-P and CMB-S4. 

In the following, we briefly describe the content of each catalog for a given frequency: 
\begin{itemize}
    \item[] \texttt{phi}, \texttt{theta}: the longitudinal and the latitudinal  coordinates of radio sources in units of radians. Since these are distributed uniformly in the sky, any system of coordinates (whether Ecliptical, Equatorial or Galactic) is equivalent; 
    \item[] \texttt{flux}:  the flux associated to each source in units of Jy; 
    \item[] \texttt{polarized flux}:  the polarized flux associated to each source in units of Jy; 
    \item[] \texttt{redshift}:  the redshift associated to each source;
    \item[] \texttt{map}:   I(QU)  map of   radio sources obtained by   projecting the (polarized) flux  entries from a given catalog  to \texttt{nside=4096} HEALPix intensity map, in units of Jy/srad;

\end{itemize}
\begin{figure}[t]
\includegraphics[trim = 0 20 0 0, clip,width=0.9\columnwidth]{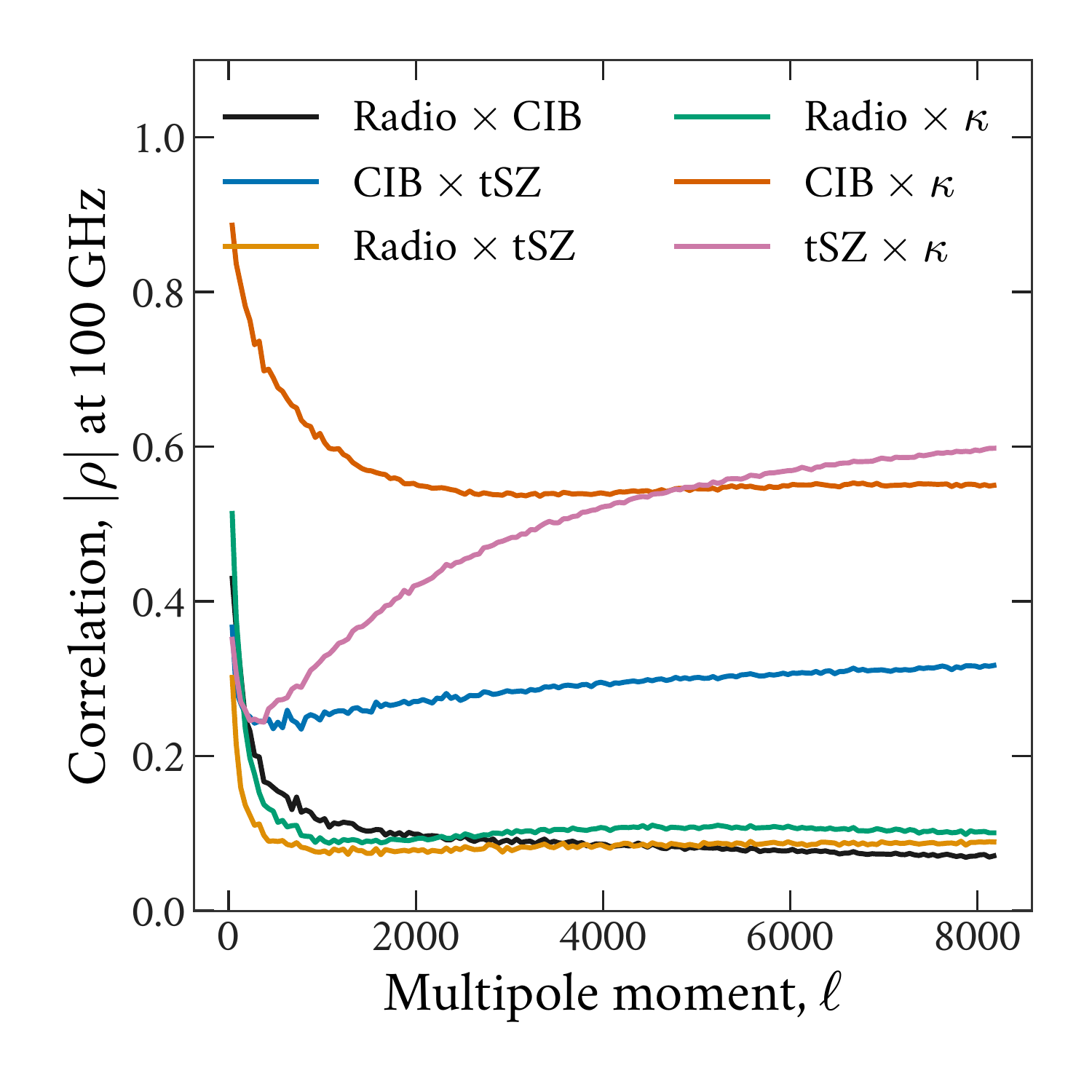}
\centering
\caption{Correlation coefficients between various Websky foreground fields at 145 GHz, defined as $\rho = C_{\ell}^{AB} / \sqrt{C_{\ell}^{AA} C_{\ell}^{BB}}$. The well-known strong correlation between CIB and tSZ is shown in blue, which is of order $\sim 30\%$. In contrast, the correlation between the radio sources with the CIB and tSZ (shown in black and orange, respectively) is of order $\sim 10\%$. }
\label{fig:corrcoeff}
\end{figure}

Scripts to reproduce the catalogs are available online on the dedicated repository\footnote{\url{https://github.com/xzackli/WebskyRadioGalaxies}}.


\section{Discussion}\label{sec:conclusions}

Using the same astrophysical modeling as in W08 and S10, we are able to reproduce the total source counts between the Fanaroff-Riley I and II morphologies as in S10, across a broad range of milimeter-wave frequencies (1.4 - 145 GHz). In Figure \ref{fig:sehgal_comparison} we compare the source counts directly, and find excellent agreement at all fluxes and frequencies. Note that instead of direct N-body simulations as used in S10, we use catalogs generated from peak patch simulations, an approximate N-body method. The simulated cosmology has also changed modestly since 2009. Despite these differences, the main discrepancies in the source counts shown in Figure \ref{fig:sehgal_comparison} arise from Poisson fluctuations. S10 used halo catalogs covering only an eighth of the sky, which they then reproduced across the full sky. Since the peak patch halo catalogs used in this work cover the full sky, they incur smaller fluctuations at bright fluxes relative to the S10 catalogs.

The astrophysical models used in W08, S10, and this work are fundamentally based on the low frequency radio luminosity function (RLF), shown in Figure \ref{fig:abundance}. Radio sources tend to be less stochastic at low frequencies, and are dominated by the steep-spectrum lobe emissions at 151 MHz. The observed 151 MHz radio luminosity function is therefore a more complete sample than those found at higher frequencies, which tend to be dominated by the highly stochastic core emissions of radio galaxies. We demonstrate that our model reproduces the observed radio luminosity functions. 

However, new data from \citet{PlanckPCCS2} released after the these models now show some discrepancies with S10. In particular, the mean and variance of the spectral indices from catalogs in S10 do not agree with those measured by \emph{Planck}. After the random sampling procedure detailed in \ref{sec:match}, we are able to reproduce the state-of-the-art of radio source observations, which we show in Figure \ref{fig:stateofar_counts}. These catalogs also model the cross-correlations between the radio galaxy sky and other observables correlated with large-scale structure, like gravitational lensing, CIB, and SZ in the Websky extragalactic suite of simulations. 

Radio galaxies also contaminate measurements of clusters in the sub-millimeter wavelength range, and we intend for our simulated sources to contribute towards analysis pipelines that are robust to this contamination. Our simulated radio galaxies exhibit this behavior, and we find that radio galaxies contribute $\sim 10$\% to the tSZ and CIB if one stacks on massive halos, since all three components inhabit the same dark matter halos. It should be noted that the source data we used for fitting the radio source model may include some confusion for dusty galaxies in the CIB component. We leave clarification of these effects to future work.

Finally, we stress that our hybrid modeling approach is phenomenological in nature, and it is representative to the extent that it is constrained by currently available date; uncertainties solely from extrapolation of the adopted fitting formula are non-negligible for fluxes below $S\sim 0.1 $ nJy. The release of new public data will help to improve the SED modelling and extrapolate the models of number counts to lower fluxes. However, the agreement with currently available data indicates that the Websky radio source  catalogs  provide simulated datasets in agreement with available data and of sufficient realism to be useful in the context of forthcoming CMB experiments in the near term. 

These simulated radio source catalogs and maps were developed as part of the Websky suite of extragalactic foreground simulations. They reproduce current data, but exhibit nonlinear structure and correlations between the radio galaxies and other tracers of the large-scale structure, such as lensing, CIB, and tSZ. We expect these simulations to be broadly useful for data analysis efforts in the millimeter, in particular for building and testing data analysis pipelines that will extract cosmology from the CMB, clusters, and lensing in the next decade. These radio source models are phenomenological and impose certain physical priors based on our understanding of AGN and galaxy evolution -- we expect that new data will shed light on these astrophysics, and that these models may need future updates from surveys in the millimeter.


\section*{Acknowledgments}

GP would like to thank Guilaine Lagache, Joacquin Gonzalez-Nuevo, Laura Bonavera, Marcella Massardi, and Vincenzo Galluzzi for useful discussions. ZL would like to thank Jo Dunkley for helpful feedback. Some of the results in this paper have been derived using the healpy~\cite{Healpix1}, HEALPix~\cite{Healpix2}, and Healpix.jl~\cite{tomasi2021} packages. This research made use of Astropy,\footnote{http://www.astropy.org} a community-developed core Python package for Astronomy \citep{astropy:2013, astropy:2018}. We also acknowledge use of the matplotlib~ \cite{Hunter:2007} package and the Python Image Library~\citep{umesh2012image} for producing plots in this paper. The Websky simulations used in this paper were developed with the continuous support of the Canadian Institute for Theoretical Astrophysics, the Canadian Institute for Advanced Research, and the Natural Sciences and Engineering Council of Canada, and were generated on the Niagara supercomputer at the SciNet HPC Consortium. SciNet is funded by: the Canada Foundation for Innovation under the auspices of Compute Canada; the Government of Ontario; Ontario Research Fund - Research Excellence; and the University of Toronto. Research at Perimeter Institute is supported
in part by the Government of Canada through the Department of Innovation, Science and Industry Canada and by the Province of Ontario through the Ministry of Colleges and Universities. 

\typeout{} 
\bibliographystyle{apsrev}
\bibliography{main}
\appendix


\end{document}